\documentclass[twocolumn,prd,superscriptaddress,showpacs]
              {revtex4}
\usepackage{graphicx,color}

\begin{document}

\title{GUP in presence of extra dimensions and lifetime of mini black holes}

\author{M.~T.~Makhviladze}
\email{mixomaxi@yahoo.com}\affiliation{Department of Physics,
Tbilisi State University, 3 Chavchavadze Ave., Tbilisi 0128,
Georgia}

\author{M.~A.~Maziashvili}
\email{maziashvili@ictsu.tsu.edu.ge}\affiliation{Department of
Physics, Tbilisi State University, 3 Chavchavadze Ave., Tbilisi
0128, Georgia} \affiliation{Institute of High Energy Physics and
Informatization, 9 University Str., Tbilisi 0186, Georgia }

\author{D.~V.~Nozadze}
\affiliation{Department of Physics, Tbilisi State University, 3
Chavchavadze Ave., Tbilisi 0128, Georgia}

\begin{abstract}
Based on the general considerations of quantum mechanics and
gravity the generalized uncertainty principle (GUP) is determined
in higher-dimensional case and on the brane, respectively. The
result is used to evaluate the effect of GUP on the dynamics of
evaporation and lifetime of mini black holes in the brane-world
models.
\end{abstract}


%04.70.-s Physics of black holes
%04.70.Dy Quantum aspects of black holes, evaporation, thermodynamics

\maketitle

\section{GUP in four dimensions}

For a quite long time since the appearance of Heisenberg
uncertainty relation in $1923$ the gravitational interaction
between photon and the particle being observed has not been
considered as it is usually assumed to be negligible. But at
increasingly large energies this interaction becomes more and more
important. The main conceptual point concerning GUP is that there
is an additional uncertainty in quantum measurement due to
gravitational interaction. We focus on different considerations of
this problem presented in \cite{Sc} and \cite{AS}. Both of these
considerations rely on classical gravitational theory. Let us give
here a brief critical review of GUP that comes from the general
arguments of quantum mechanics and gravity. We set $\hbar=c=k_B=1$
in what follows. The idea based on micro-black hole
\emph{gedanken} experiment goes as follows \cite{Sc}. As it was
derived by Heisenberg the uncertainty in the position of the
electron when it interacts with the photon is given by $\Delta
x\Delta p \geq 1/2$ \cite{LL}. By taking in the high energy
situation $\Delta E\approx \Delta p$ the Heisenberg relation takes
the form $\Delta E\Delta x \geq1/2$. So that to measure the
position of electron more precisely the energy of photon should be
increased. But this procedure is limited because there is the
gravitational radius associated with $\Delta E$, $r_g=2G_4\Delta
E$, for which the $2r_g$ becomes greater than $\Delta x$ for
$\Delta E > 0.35\,G_4^{-1/2}$ and therefore the region where
$\Delta E$ is located becomes hidden by the event horizon. ($G_4$
is the four-dimensional Newton's constant). Based on this
discussion the combination of the above inequalities gives
\begin{equation}\label{genunp}\Delta x\geq \left\{
\begin{array}{ll}1/2\Delta E ~~~~~~~~~ \mbox{if}~~~~ \Delta E\leq 0.35
G_4^{-1/2}
\\4G_4\Delta E~~~~~~~~\mbox{if}~~~~ \Delta E > 0.35
G_4^{-1/2} \end{array}\right.~.\end{equation} The
Eq.(\ref{genunp}) implies a minimal attainable uncertainty in
position $\Delta x_{min}=1.43G_4^{1/2}$. While in ordinary quantum
mechanics $\Delta x$ can be made arbitrarily small by letting
$\Delta E$ grow correspondingly. Combining the Eq.(\ref{genunp})
into a single one in the linear way one gets the expression
similar to what was obtained previously in the string theory
framework \cite{Ve}
\begin{equation}\label{lcgup}\Delta x\geq {1\over 2\Delta
E}+4G_4\Delta E~.\end{equation} The approach proposed in \cite{AS}
is to calculate the displacement of electron caused by the
gravitational interaction with the photon and add it to the
position uncertainty. The photon due to gravitational interaction
imparts to electron the acceleration given by $a=G_4\Delta E/r^2$.
Assuming $r_0$ is the size of the interaction region the variation
of the velocity of the electron is given by $\Delta v \sim
G_4\Delta E/r_0$ and correspondingly $\Delta x_g \sim G_4\Delta
E$. Therefore the total uncertainty in the position is given by
\begin{equation}\label{genunpn}\Delta x\geq {1\over 2\Delta
E}+G_4\Delta E~.\end{equation}

As one sees the mechanisms proposed in \cite{Sc} and \cite{AS}
giving rise to the gravitational uncertainty are quite distinct
though they provide qualitatively the same gravitational
uncertainty term which in general has to be taken with some
numerical factor of order unity \cite{Ve}. At a first glance they
complement one another in that for energies $\Delta E\leq
0.4G_4^{-1/2}$ the GUP is given by Eq.(\ref{genunpn}) while for
$\Delta E > 0.4G_4^{-1/2}$ one can take the second mechanism
Eq.(\ref{lcgup}) as it was proposed in \cite{Sc}. But to be more
precise the collapse of $\Delta E$ puts simply the limitation on
the measurement procedure. So in principle it is no longer
conceivable for $\Delta E > 0.4G_4^{-1/2}$ to proceed the
measurement. The minimum position uncertainty then becomes $\Delta
x_{min}=1.63G_4^{1/2}$.

\section{GUP in higher dimensional case}

Let us make a straightforward generalization of the ideas given in
the preceding section to a higher dimensional case. The $D=4+n$
dimensional Black hole has the form \cite{TMP}

\begin{equation}\label{hdschs}ds^2 = - h(r)\,dt^2 + \frac{dr^2}{h(r)} + r^2
d\Omega_{2+n}^2, \end{equation} with \[h(r) =
1-\frac{\mu}{r^{n+1}}~,~~~~\mu=\frac{16\pi G_D
M\Gamma[(n+3)/2]}{(n+2)\,2 \pi^{(n+3)/2}}~,\] where the parameter
$G_D$ stands for the $D$-dimensional Newton's constant and
$\Omega_{2+n}$ denotes the volume of a $2+n$ dimensional unit
sphere. The higher dimensional Newton's law has the form

\begin{equation}\label{exnew}F=G_D\frac{m_1m_2}{r^{2+n}}~.\end{equation}

\begin{table}[t]
\begin{tabular}{|c||l|l|l|l|}
\hline ~~n~~&~~~~$\Delta E_1$~~~~&~~~~$\Delta
E_2$~~~~~&~~Length~&~~~~Mass~\\ \hline $2$ &~0.715307~&~162.088~
&~0.699001~&~0.0894133~\\ $3$ &~0.719459~&~68.1812~
&~0.694967~&~0.0449661~\\ $4$ &~0.705562~&~43.7701~
&~0.708655~&~0.0220487~\\$5$ &~0.684841~&~33.3681~&~0.730096~&~0.0107006~\\
$6$ &~0.662058~&~27.758221~&~0.755221~&~0.0051723~
\\$7$ &~0.639293~&~24.29000~&~0.782114~&~0.0024972~
\\\hline\end{tabular}

\caption{The region where the collapse of $\Delta E$ occurs is
given by ($\Delta E_1,~\Delta E_2$). "Length" denotes the minimal
position uncertainty and "Mass" stands for the minimal black hole
mass that follows from the minimal observable length. All
quantities given here are written in the Planck units.}
\end{table}

Correspondingly the electron will experience an acceleration due
to gravity
\[a=G_D{\Delta E\over r^{2+n}}~.\] As it follows from the standard quantum mechanical uncertainty relations the characteristic
time and length scales for interaction are given by $\Delta
E^{-1}$ when the energy $\Delta E$ is used for the measurement
\cite{LL}. Then the variation of velocity and the displacement of
electron are given by
\[\Delta v \sim G_D\Delta E^{2+n}~,~~~~\Delta x_g \sim G_D\Delta
E^{1+n}~.\] The derivation of gravitational uncertainty term is
somewhat rough in that, in general, it may contain some numerical
multiplier, which for the stringy induced GUP is of order unity
\cite{Ve}. Considering simply the linear combination, the total
uncertainty in the position takes the form
\begin{equation}\label{exdgup}\Delta x\geq \frac{1}{2\Delta E}+\beta G_D\Delta E^{1+n}~,\end{equation}
where $\beta $ is some numerical factor mentioned above. The
Eq.(\ref{exdgup}) predicts the minimum position uncertainty
\begin{equation}\label{minobdis}\Delta
x_{min}=l_p\beta^{\frac{1}{2+n}}\left(\frac{1}{2}(2+2n)^{\frac{1}{2+n}}+(2+2n)^{-\frac{1+n}{2+n}}\right)~,\end{equation}
which takes place for \[\Delta E_{cr}=\frac{m_p}{(2\beta+2\beta
n)^{1/(2+n)}}~,\] where $l_p=m_p^{-1}=G_D^{1/(2+n)}$. But as we
remember from the preceding section further increase of energy can
result the collapse of $\Delta E$ if two times the gravitational
radius exceeds the $\Delta x$. In this case it was suggested in
\cite{Sc} to take the uncertainty term as two times the
gravitational radius
\begin{equation}\label{uncergrrad}\Delta x \geq \frac{1}{2\Delta E}+2\left(\frac{16\pi G_D
\Gamma[(n+3)/2]\Delta E}{(n+2)\,2
\pi^{(n+3)/2}}\right)^{1/(1+n)}~.\end{equation} Thus, in higher
dimensional case the two approaches \cite{Sc},~\cite{AS} give
qualitatively different answers in contrast to the four
dimensional one. One can see that the careful consideration gives
the higher-dimensional GUP different from that one obtained
previously in \cite{SC} where the Eq.(\ref{uncergrrad}) was
assumed for all values of $\Delta E$.

\section{GUP on the brane}

Let us not to go into details of the extra-dimensional models
\cite{ADDGRS} but merely recall a few common features relevant for
our consideration. The common features are a low fundamental
Planck scale $m_p\sim$TeV, the localization of the standard model
particles on the brane and propagation of gravity throughout the
higher dimensional space. There is a length scale $L$, much
greater than TeV$^{-1}$, beneath of which the gravitational
interaction has the form Eq.(\ref{exnew}) while beyond this scale
we have the standard four-dimensional law. In what follows we
restrict ourselves to the ADD model.

From the above discussion one gets the following expression for
GUP

\begin{equation}\label{genonb}\Delta x\geq \left\{
\begin{array}{ll}1/2\Delta E+\beta G_D\Delta E^{1+n} ~~~ \mbox{if}~~~~ \Delta E^{-1}\leq
L
\\1/2\Delta E+\alpha G_4\Delta E~~~~~~~~\mbox{if}~~~~ \Delta E^{-1} > L \end{array}\right.~,\end{equation}
where $\alpha$ is of order unity \cite{Ve} and $\beta$ has to be
determined by the junction of these inequalities at $\Delta
E=L^{-1}$. In this way one gets \begin{equation}\beta = \alpha
\frac{G_4}{G_D}L^n~.\end{equation} As it was mentioned in the
first section, the GUP obtained in the framework of string theory
coincides with a model-independent derivation based on the general
considerations of quantum mechanics and gravity. However, as it is
shown in section two, the latter approach gives the
higher-dimensional GUP different from that one obtained in
four-dimensional case and GUP on the brane incorporates both of
them Eq.(\ref{genonb}). For the ADD model the size and number of
extra dimensions are related as follows $L\sim 10^{-17+30/n}$cm.
So in this case one finds $\beta\sim 10^{-4-n}$ and $\Delta
x_{min}\sim 10^{-(4+n)/(2+n)}l_p$ as it follows from
Eq.(\ref{minobdis}). However, the minimal observable distance is
determined rather due to collapse of $\Delta E$ than merely by the
Eq.(\ref{minobdis}). It is natural because the collapse of $\Delta
E$ puts simply the bound on the measurement procedure. In this way
for minimal observable lengths and the black hole remnant masses
one finds the values represented in Table I.

\section{Brane-localized black hole emission}

It is widely believed that if the fundamental scale of gravity
actually lies in the TeV range a very spectacular prediction can
be made about the creation of small black holes at colliders or in
high-energy cosmic ray interactions \cite{DLGTFS}. The size of
black hole produced in this way is typically assumed to be much
smaller than the characteristic length of extra dimensions. Then
it seems reasonable to describe such objects by higher dimensional
Schwarzschild solution Eq.(\ref{hdschs}).

\begin{figure}[t]
\includegraphics[width=\linewidth]{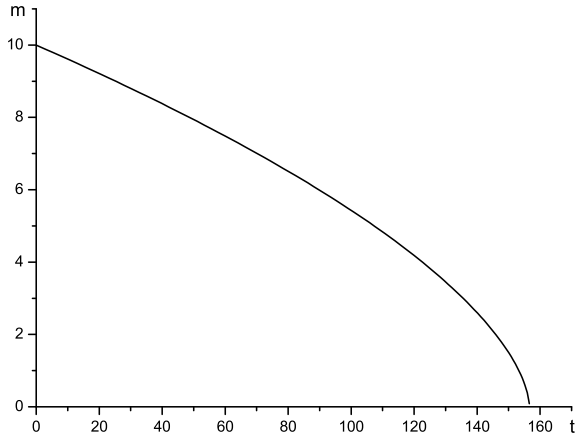}\\
\caption[lifetime]{ADD $n=2$ model: The mass of the black hole
versus time in Planck units.}
\end{figure}

Once produced, the black hole would decay very rapidly to a
spectrum of particles by Hawking radiation. For mass shedding
formula in $D$ dimensional space-time, under assumption that the
black hole emits mainly the massless particles we have the
following expression \cite{EHM, GCL}

\begin{equation}\frac{dM}{dt}\approx -7.4931\,r_D^2T^4-g_{bulk}\sigma_DA_DT^D~,\end{equation}
where
\[r_D=\left(\mu\frac{D-1}{2}\right)^{\frac{1}{D-3}}\left(\frac{D-1}{D-3}\right)^{\frac{1}{2}}~,~A_D=\Omega_{D-2}r_D^{D-2}~,\]
and $\sigma_D$ denotes the $D$ dimensional Stefan-Boltzmann
constant
\[\sigma_D=\frac{\Omega_{D-2}}{4(2\pi)^{D-1}}\Gamma(D)\zeta(D)~.\]
Since only gravity is allowed to propagate in the bulk, $g_{bulk}$
simply counts the number of polarization states of the graviton,
namely
\[g_{bulk}=\frac{D(D-3)}{2}~.\]

In the framework of heuristic approach \cite{ACS,CD} to the
Hawking radiation the black hole is envisioned as a cube with size
two times the gravitational radius, inner space of which is
inaccessible for outer observable, and the characteristic energy
of the emitted particles is estimated with the use of GUP. Then
the lower value of energy obtained in this way is identified with
the radiation temperature
\[T=\frac{D-3}{\pi}\Delta E~.\] The main observation achieved in the framework of
this approach is that at the Planck scale black hole ceases to
radiate, even though its temperature reaches a maximum. It cannot
radiate further and becomes an inert remnant of about Planck mass.

The problem of brane-localized mini black hole evaporation was
studied in the framework of stringy induced GUP in \cite{CD}. Here
we focus on the GUP derived in the preceding section. From the
expression of GUP derived in section three one sees that its
effect on the Hawking temperature becomes negligibly small for the
ADD model. The Figure 1 shows the mass decay of the black hole
with initial mass $M=10 m_p$ for ADD model with $n=2$. The
corresponding lifetime is estimated as $\sim 156t_p$. Without GUP
one obtains practically the same result.

\section{Conclusion}

Following to the papers \cite{Sc,AS} we have defined the GUP on
the brane which is further used for evaluating of evaporation of
brane-localized mini black holes in the framework of heuristic
approach given in \cite{ACS}. As it is shown for ADD model the GUP
effect on the black hole evaporation is strongly suppressed
because of small $\beta$ factor. We have also evaluated the masses
of the black hole remnants the existence of which increase the
lower cutoff for the black hole production reducing therefore the
rate of corresponding events.

In regard with the GUP approach to the black hole evaporation a
few remarks are in order. In general, GUP assumes two values of
$\Delta E$ for a given $\Delta x$. In four-dimensional case the
situation may be somewhat simplified by omitting the branch
$\Delta E > 0.4G_4^{-1/2}$ in GUP as it corresponds to the
collapse of $\Delta E$ (see section one). In this regard the
situation in presence of extra dimensions is more complicated
because the region of collapse of $\Delta E$, given by
$\left(\Delta E_1,~\Delta E_2\right)$, is bounded. Only the lower
value of $\Delta E$ is applicable to the black hole radiation for
it gives the correct asymptotic dependence of the Hawking
temperature on the black hole mass. The question naturally arises
is why the nature selects the lower one in the case of black hole
emission whereas in the framework of GUP both of the solutions
have the same right of existence. So we face the challenge of
understanding what is the status of that part of the GUP
corresponding to the higher values of $\Delta E$. Does the nature
uses the higher value of $\Delta E$ instead of the lower one in
some cases? If so, what is the intrinsic principle that selects
which solution of GUP should be used for a given process? Also it
is necessary to know what are the properties of remnants left
behind the black hole evaporation in the GUP approach. Another
important point is the quantum gravity effects since GUP takes
into account the gravitational interactions at the Planck scale
\cite{BR}. Moreover, the black hole remnant may be rather due to
quantum gravitational effects (the nature of which is well
established in this case \cite{BR}) than the GUP \cite{Ma}.
(Certainly, a big obstacle so far is the lack of a direct
experimental hint that there is a need for a quantum theory of
gravity).

\begin{acknowledgments}
The authors are indebted to the participants of seminar at the
Chair of Theoretical Physics, Tbilisi State University, for useful
discussions. The work of M.\,A.\,M. was supported by the
\emph{Georgian President Fellowship for Young Scientists} and the
grant FEL. REG. $980767$.
\end{acknowledgments}

%%%%%%%%%%%%%%%%%%%%%%%%%%%%%%%%%%%%%%%%%%%%%%%%%%%%%%%%%%%%%%%%%%%%%%%%

\end{document}